\documentclass[aps,prl,twocolumn,amsmath,amssymb,superscriptaddress,floatfix]{revtex4-1}

\usepackage{graphicx, multirow, xcolor, bm}
\usepackage{amsfonts,amssymb,amsmath}
\usepackage{graphicx,dcolumn,bm,color,braket,slashed}
\usepackage{times, comment, mathtools, extarrows,textcomp}

\begin{document}

\title{
Topological Domain-Wall States Hosting Quantized Polarization and Majorana Zero Modes
Without Bulk Boundary Correspondence
}

\author{Sang-Hoon Han}
\thanks{These authors contributed equally to this work.}
\affiliation{Department of Physics, Hanyang University, Seoul 04763, South Korea}

\author{Myungjun Kang}
\thanks{These authors contributed equally to this work.}
\affiliation{Department of Physics, Hanyang University, Seoul 04763, South Korea}

\author{Moon Jip Park}
\email{moonjippark@hanyang.ac.kr}
\affiliation{Department of Physics, Hanyang University, Seoul 04763, South Korea}

\author{Sangmo Cheon}
\email{sangmocheon@hanyang.ac.kr}
\affiliation{Department of Physics, Hanyang University, Seoul 04763, South Korea}
\affiliation{Research Institute for Natural Science and High Pressure, Hanyang University, Seoul 04763, South Korea}

\begin{abstract}
Bulk-boundary correspondence is a concept for topological insulators and superconductors that determines the existence of topological boundary states within the tenfold classification table.
Contrary to this belief, we demonstrate that topological domain-wall states can emerge in all forbidden 1D classes in the classification table using representative generalized Su-Schrieffer-Heeger and Kitaev models, which manifests as quantized electric dipole moments and Majorana zero modes, respectively.
We first show that a zero-energy domain-wall state can possess a quantized polarization, even if the polarization of individual domains is not inherently quantized.
A quantized Berry phase difference between the domains confirms the non-trivial nature of the domain-wall states, implying a general-bulk-boundary principle, further confirmed by the tight-binding, topological field, and low-energy effective theories.
Our methodology is then extended to a superconducting system, resulting in Majorana zero modes on the domain wall of a generalized Kitaev model.
Finally, we suggest potential systems where our results may be realized, spanning from condensed matter to optical.
\end{abstract}

\maketitle


The tenfold classification of the topological periodic table provides a systematic understanding of topological insulators and superconductors in the presence of time-reversal ($T$), particle-hole ($C$), and chiral ($\Gamma$) symmetries.~\cite{schnyder2008classification, chiu2016classification}.
The bulk-boundary correspondence~\cite{qi2011topological, hasan2010colloquium}, as a guiding principle of topological materials, predicts robust topological edge/surface states against perturbation, which have potential applications
in many subfields of physics, including spintronics
\cite{he2022topological, tokura2019magnetic}, 
ultracold atomic gases
\cite{atala2013direct, cooper2019topological}, 
quantum information
\cite{nayak2008, stern2013topological}, 
photonics
\cite{meier2016, ozawa2019},
and mechanics
\cite{zhou2017, zeng2021}.
New types of topological phases have been discovered using finer topological classifications
in topological crystalline insulators/superconductors, a crystalline point group symmetry protects topological boundary states
\cite{shiozaki2014topology,cornfeld2019classification}.
Similarly, higher-order topological phases~\cite{khalaf2018symmetry}---gapped bulk bands and gapless boundary states with codimension greater than one---and topological semimetal phases are also classified ~\cite{yang2014classification,armitage2018weyl}. 

Circumventing the usual classifications, several studies have tried to find a method that still results in topological phases; for example, sub-symmetry-protected topological phases and quasi-symmetry-protected topological semimetal were investigated.
These classifications lead to robust topological applications as well as unexpected topology beyond the usual space group classifications even in the absence of full symmetry~\cite{wang2023sub,guo2022quasi}.
As an alternative guiding principle, our endeavors are focused on revealing whether zero-energy topological domain-wall states can exist for systems of topologically trivial cases, as shown in Fig.~1.

\begin{figure}[t]
\includegraphics[width=0.9\linewidth]{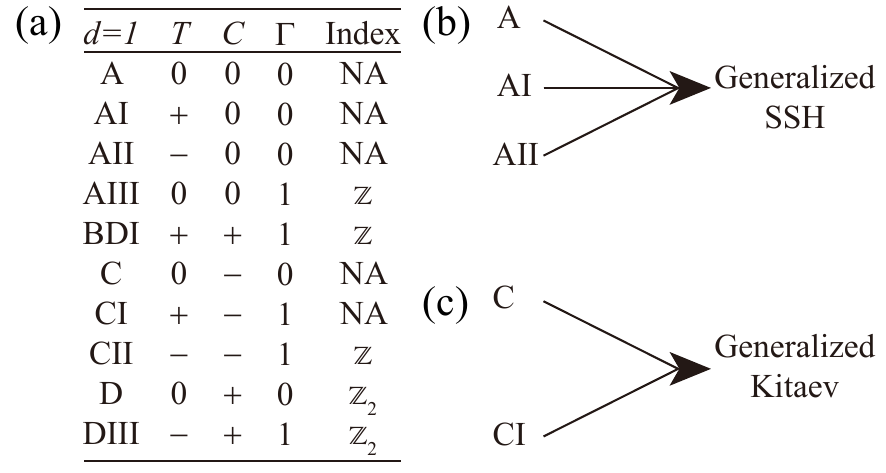}
\caption{\label{fig.1.aztable} 
\textbf{(a)} Classification of the tenfold Altland-Zirnbauer classes in 1D.
The lack of symmetry is presented with $0$.
For the antiunitary time-reversal ($T$) and particle-hole ($C$) symmetries, $+$ and $-$ distinguish the cases in which the symmetry operators' squares become $I$  or $-I$.
$\mathbb{Z}$ and $\mathbb{Z}_2$ represent topological indices, while NA indicates topology is not allowed.
\textbf{(b)} Schematics of 1D non-topological classes simplifying to the generalized Su-Schrieffer-Heeger (SSH) and
\textbf{(c)} Kitaev models.
}
\end{figure}

Generally, non-trivial topological indices are determined by comparing them with the invariant in a vacuum (atomic limit). 
Here, even in a trivial bulk, we show that the finer classification can exist where the difference in the topological indices of the two trivial bulks is still quantized.
Our result differs from the topological classification for defects~\cite{PhysRevB.82.115120}
since the defect classifications demand that one of the domains is still topological.
We start our discussion by demonstrating the domain walls of insulating chains of the AI class.
While any one of the domains does not have a quantized polarization without symmetry protection, we find the quantized difference of the Berry/Zak phase between the adjoined domains, resulting in the zero-energy domain-wall state.
This concept is expanded to other classes, summarized in Fig.1(b,c); the A, AI, and AII classes can be simplified into a generalized Su-Schrieffer-Heeger (SSH) model for insulating systems [Fig.~\ref{fig.1.aztable}(b)] and the C and CI classes to a generalized Kitaev model for superconducting systems [Fig.~\ref{fig.1.aztable}(c)].
Consequently, such insulating and superconducting systems demonstrate quantized electric polarization and Majorana zero modes, respectively.


\begin{figure}[t]
\includegraphics[width=0.9\linewidth]{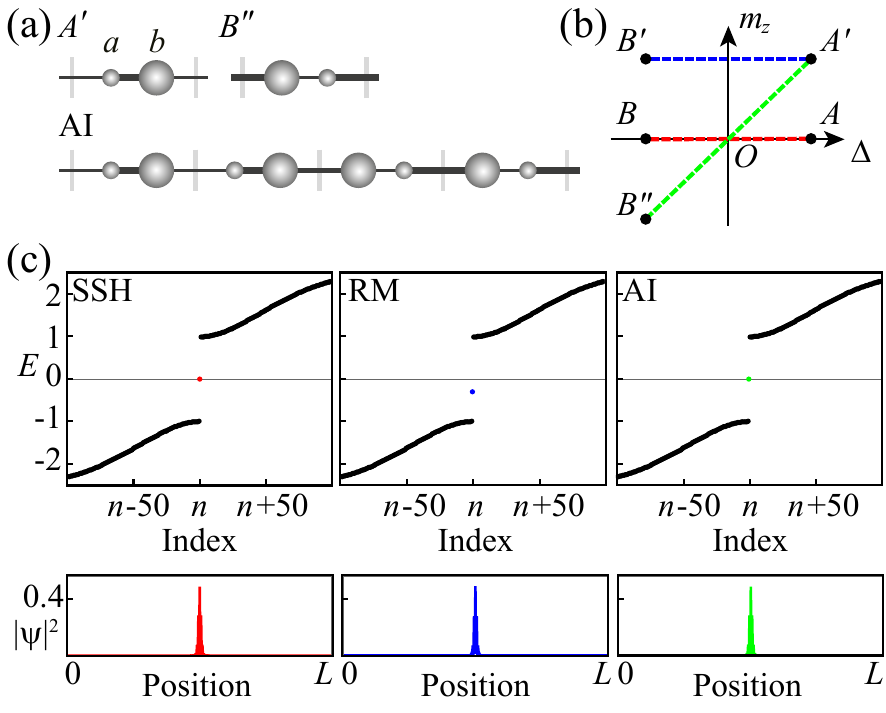}
\caption{\label{fig.2.GSSmodel} 
Generalized Su-Schrieffer-Heeger (GSSH) model and various domain walls.
\textbf{(a)} Schematics for two groundstates and a domain-wall configuration for the AI class.
$a$ and $b$ in the unit cell represent two atoms.
\textbf{(b)} Order-parameter space comprised of the energy-valued dimerization and onsite energy.
The red, blue, and green dotted lines indicate the SSH, Rice-Mele (RM), and AI domain-wall configurations.
\textbf{(c)} Energy eigenvalues of finite SSH, RM, AI domain walls, and the wavefunctions of the ingap states.
The parameters are given in Ref.~\cite{paraSSH}.
Note that zero-energy states emerge even for $\xi_1\neq\xi_2$ [Fig.~S2].
}
\end{figure}

\textit{Generalized bulk-boundary correspondence in insulating systems}---We first investigate the minimal model of the AI class as the representative model among A, AI, and AII classes \cite{Supple}.
The Hamiltonian of the generalized Su-Schrieffer-Heeger model (GSSH) is comprised of the Hamiltonian of the original SSH model $H_{\text{SSH}}$~\cite{su_solitons_1979} and the symmetry-breaking potential terms $H_\text{on}$~\cite{rice1982} [Fig.~\ref{fig.2.GSSmodel}(a)]:
\begin{align*}
    H_\text{AI} &= H_{\text{SSH}}+H_\text{on},\\
    H_{\text{SSH}} &=\sum_n t_{n+1,n}c^\dagger_{n+1}c_{n}+h.c.,\\
    H_\text{on}&=\sum_n m_z\left(c^\dagger_{2n-1}c_{2n-1}-c^\dagger_{2n}c_{2n}\right),
\end{align*}
where $c_n^\dagger$/$c_n$ indicates the creation/annihilation operator for site $n$. The nearest-neighbor hopping parameter is $t_{n+1,n}=t_0+(-1)^n\Delta$ with $t_0$ and $\Delta$ being the hopping amplitude and energy-valued dimerization, respectively.
$m_z$ is the onsite energy.
The corresponding Bloch Hamiltonian is given as,
\begin{align}
H_\text{AI}(k) = 2t_0\cos{k}\sigma_x-\Delta\sin{k}\sigma_y+m_z\sigma_z,\label{RMHam}
\end{align}
where $\sigma_i$ indicates the $i$-th Pauli matrix.
The system is time-reversal symmetric ($T=K$).
In particular, $\Delta$ and $m_z$ act as order parameters indicating various  groundstates [Fig.~\ref{fig.2.GSSmodel}(a,b)].
The energy eigenvalue is $E=\pm\sqrt{4t_0^2 \cos^2 k+\Delta^2\sin^2 k+m_z^2}$.
In the domain-wall configuration, $\Delta$ and $m_z$ are spatially varying functions interpolating energetically degenerate groundstates [Fig.~\ref{fig.2.GSSmodel}(b)].
For instance, $ m_z=0$ ($ m_z = \text{const.}$) throughout the chain corresponds to the SSH (Rice-Mele or RM) domain wall with $\Delta(x) = \Delta_0 \tanh{(x/\xi_1)}$~\cite{su_solitons_1979,rice1982}.
When $m_z$ is interpolated such as $m_z(x)=m_{z,0}\tanh{(x/\xi_2)}$, a new type of domain-wall state emerges connecting the $A'$ and $B''$ groundstates [Fig.~\ref{fig.2.GSSmodel}(b)], labeled as the AI domain wall.
Here, $\xi_i$ is the characteristic length scale of the domain wall.
The energy eigenvalues of the three domain-wall configurations are shown in Fig.~\ref{fig.2.GSSmodel}(c);
the SSH (RM) domain-wall state is located at zero (non-zero).
Unexpectedly, the AI domain wall exhibits the emergence of the zero-energy state.
Only the SSH and AI domain walls have global chiral symmetry, $\left\{H_\text{AI},\Gamma\right\}=0$, and this symmetry protects the zero-energy states~\cite{Supple}.

\begin{figure}[t]
\includegraphics[width=0.95\linewidth]{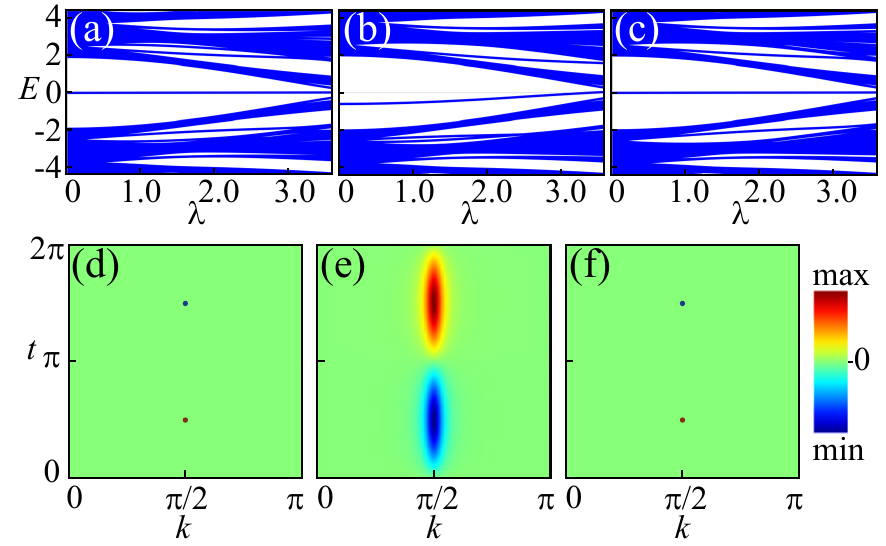}
\caption{\label{fig.3.GSSdisorder}
The energy eigenvalues and Berry curvatures for the \textbf{(a,d)} SSH, \textbf{(b,e)} RM, and \textbf{(c,f)} AI domain walls under quasi-periodic disorder.
The disorder at site $i$ is $\left(-1\right)^i \lambda \cos\left(2\pi\beta i\right)$ with disorder strength $\lambda$ and the inverse golden ratio $\beta$.
The parameters are the same as in Fig~\ref{fig.2.GSSmodel}.
}
\end{figure}

To see the robustness of the chains' zero-energy state, the energy spectra are calculated in the presence of the onsite staggered quasi-periodic disorders for the three types of domain walls [Fig.~\ref{fig.3.GSSdisorder}(a-c)].
Such disorders are chosen as they are simple to enforce, yet their descriptiveness is powerful enough to give sharp results, such as reentering the topological phase~\cite{roy2021reentrant,roy2023critical}.
Figure~\ref{fig.3.GSSdisorder}(a-c) shows that the robustness of the zero-energy state of the AI domain wall to disorder is comparable to that of the SSH domain wall, while the RM domain wall fluctuates easily under the same disorder.

To understand the origin of the bound state, we examine the low-energy Dirac-type effective Hamiltonian for the GSSH model via the Jackiw-Rebbi method~\cite{jackiw1976}, confirming the condition for the domain-wall state being stable.
Taking the Dirac approximation at $k=-\frac{\pi}{2}+k'$, the effective Hamiltonian is obtained as $H_\text{eff}=2t_0 k' \sigma_x-\Delta(x) \sigma_y+m_z(x) \sigma_z$ with $k'=-i\partial_x$ in the continuum limit.
Hence, the zero-energy domain-wall solution will be $\psi(x)=\mathcal{N}\exp\left[\mp\frac{1}{2t_0}\int^xdx'\sqrt{\Delta^2+m_z^2}\right]\left(im_z,\Delta \pm \sqrt{\Delta^2+m_z^2}\right)^T$, 
where $\mathcal{N}$ is the normalization factor, $\Delta$ and $m_z$ are naturally $x$-dependent functions, and the double signs are in the same order with the upper (lower) signs for $x>0$ ($x<0$).
Moreover, the SU(2) unitary transformation via a unitary operator $U=\exp\left[-\frac{i}{2}\tan^{-1}\left(\frac{m_z(x)}{\Delta(x)}\right)\sigma_x\right]$ transforms the effective Hamiltonian into that of the SSH model:
\begin{align*}
H_\text{eff}'
=2t_0 k' \sigma_x-\text{sgn}(\Delta) \sqrt{\Delta^2+m_z^2}\sigma_y+\partial_x f \sigma_x,
\end{align*}
where $\Delta$ and $m_z$ are $x$-dependent functions, and $f=-t_0 \tan^{-1}\left(\frac{m_z(x)}{\Delta(x)}\right)$ term acts as a small oscillatory correction term~\cite{Supple}; in the case of $\xi_1=\xi_2$, the $f$ term can be ignored as $\partial_x f=0$.
Therefore, within the low-energy effective theory, the GSSH model can be equivalently transformed into the BDI class and, therefore, can have zero-energy topological modes.
Such analytical result also highlights the restriction for the emergence of the zero-energy domain-wall state; $\Delta(x)$ and $m_z(x)$ are interpolated such that there exists a so-called topological phase transition point where $\sqrt{\Delta(x)^2+m_z(x)^2}$ is zero, similar to the SSH model~\cite{nt3}.

We can understand the origin of the zero-energy domain-wall states and its relation to polarization by considering the generalized bulk-boundary correspondence based on the Berry/Zak phase~\cite{berry1984quantal,Zakphase1989,vanderbilt2018berry}.
The Berry curvature is defined in 2D space-time, with the additional auxiliary time axis, i.e., the system is adiabatically transformed from one to another groundstate along the lines seen in Fig.~\ref{fig.2.GSSmodel}(b) with respect to time, equivalent to the case where a domain wall moves slowly~\cite{Supple}.
With this definition, the Berry phase difference between energetically degenerate but topologically distinct groundstates is calculated from the Berry curvature by applying Stoke's theorem [Table~S1].
The resulting Berry phase difference between the phases is found to be quantized to $\pi$ (arbitrary) for both the SSH and AI (RM) domain walls, indicating a topological (trivial) nature [Fig.~\ref{fig.3.GSSdisorder}(d-f)].
This indicates that even for seemingly trivial cases, zero-energy domain-wall states can still emerge via our general bulk-boundary correspondence; 
the quantized topological quantity between two adjoined domains protects the zero-energy interface state.

\begin{figure}[t]
\includegraphics[width=0.9\linewidth]{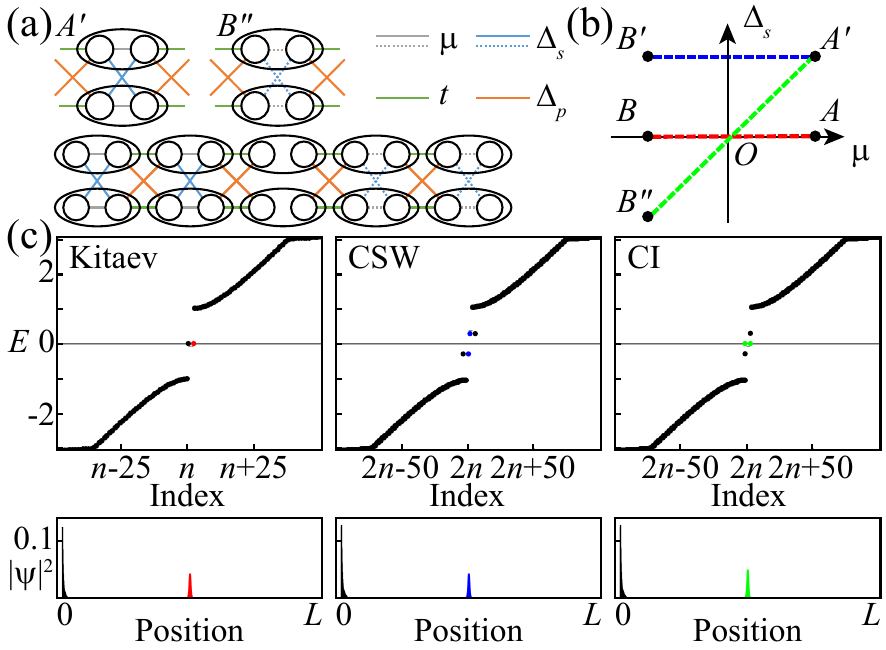}
\caption{\label{fig.4.GKmodel}
Generalized Kitaev (GK) model and various domain walls.
\textbf{(a)} Schematics for two groundstates and a domain-wall configuration for the CI class in the Majorana fermion picture,
where the fermion operator on each physical site (oval) is split up into two Majorana operators (circle).
The grey, green, blue, and orange lines indicate chemical potential, hopping, $s$- and $p$-wave pairings, respectively.
The solid and dotted blue (grey) lines indicate $\Delta_s>0$ and $\Delta_s<0$ ($|\mu|<|2t|$ and $|\mu|>|2t|$) with the empty part being $\Delta_s=0$ ($|\mu|=|2t|$).
\textbf{(b)} Order-parameter space composed of the chemical potential and $s$-wave pairing function.
The red, blue, and green dotted lines indicate the Kitaev, constant $s$-wave (CSW), and CI domain walls, and the origin represents $O=(-2t,0)$.
\textbf{(c)} Energy eigenvalues and the wavefunctions of the domain-wall (edge) states, colored red, blue, and green (black).
The parameters are given in Ref.~\cite{KT}.
}
\end{figure}

\textit{Generalization to superconducting systems}---We now focus on the minimal spinless model of the CI class as the representative model between the C and CI classes, which is denoted as the generalized Kitaev (GK) model.
Unlike the conventional Kitaev model utilizing $p$-wave pairing potential only~\cite{kitaev2001unpaired}, we consider both $s$- and $p$-wave pairing potentials to examine the existence of the Majorana zero modes in the GK model.
Thus, the model Hamiltonian $H_\text{CI}$ is comprised of three parts $H_n$, $H_s$, and $H_p$ which are normal state, $s$- and $p$-wave pairing Hamiltonians [Fig.~\ref{fig.4.GKmodel}(a)]:
\begin{align*}
	H_\text{CI}&=H_n+H_s+H_p,\\
	H_n&=\sum_{\gamma,j} -t \left(c_{\gamma,j}^\dagger c_{\gamma,j+1}+c_{\gamma,j+1}^\dagger c_{\gamma,j}\right)-\mu c_{\gamma,j}^\dagger c_{\gamma,j},\\
	H_s&=\sum_j \Delta_s \left(-c_{\alpha,j} c_{\beta,j}+c_{\beta,j} c_{\alpha,j}\right) + h.c.,\\
	H_p&=\sum_{\gamma,j} \Delta_p\left(c_{\gamma,j+1} c_{\gamma+1,j}-c_{\gamma,j} c_{\gamma+1,j+1}\right) + h.c.,
\end{align*}
where $c_{\gamma,j}^\dagger$/$c_{\gamma,j}$ indicates the creation/annihilation operator for orbital $\gamma=\alpha, \beta$ at site $j$.
Here, $\mu,t,\Delta_s,$ and $\Delta_p$ are the chemical potential, nearest-neighbor hopping parameter, and $s$- and $p$-wave pairing gaps, respectively.
Thus, the Bogliubov-de-Gennes (BdG) Hamiltonian is given by
\begin{align}
    H_\text{CI}(k) =-\left(2t\cos{k}+\mu\right)\tau_z+\Delta_s\tau_y\sigma_y+2\Delta_p\sin{k}\tau_y\sigma_x,
    \label{CIHam}
\end{align}
where the Pauli matrices $\tau_i$ and $\sigma_j$ indicate the Nambu space and orbital degrees of freedom. 
The system has time-reversal ($T=K$), particle-hole ($C=\tau_y\sigma_zK$), and chiral ($\Gamma=\tau_y\sigma_z$) symmetries.
Unlike the Kitaev model, the simplest representation of the CI class is a $4\times4$ matrix Hamiltonian due to the Fermi statistics, $\Delta(k)^T=-\Delta(-k)$, and symmetry constraints of the CI class, $CH(k)C^{-1}=-H(-k)$, which forbids the presence of $s$- and $p$-wave pairings for a 1D chain within a $2\times2$ BdG Hamiltonian.

We now discuss the degenerate groundstates and the domain-wall state connecting them.
For simplicity, we focus on the physics near the Fermi level, where the superconducting gap opens near $k=0$, similar to the Kitaev model.
Moreover, we take a limit of large $\Delta_p$ where the $p$-wave dispersion is more dominant than the electronic dispersion $2t \cos k$, i.e., a degenerate limit.
Then, the energy eigenvalue of Eq.~(\ref{CIHam}) is given by
$E=\pm\sqrt{\bar \mu^2+\Delta_s^2+4\Delta_p^2\sin^2{k}}$.
Hence, $\bar \mu = \mu+2t$ and $\Delta_s$ act as our order parameters of the system with a fixed $\Delta_p$ [Fig.~\ref{fig.4.GKmodel}(b)], and thus the spatial functions, $\bar \mu(x)$ and $\Delta_s(x)$, result in a domain-wall configuration.
The schematic of Fig.~\ref{fig.4.GKmodel}(a) shows a domain-wall configuration in the Majorana fermion representation, which effectively shows the localized Majorana domain-wall mode compared to the electron-hole representation in Fig.~S3(a).
For instance, if $\Delta_s(x)=0$ throughout the chain, the GK model reduces to a two-orbital Kitaev domain wall with $\bar\mu(x) = \mu_0\tanh{(x/\xi)}$, connecting the $A$ and $B$ groundstates [Fig.~\ref{fig.4.GKmodel}(b)], here, $\xi$ is the characteristic length of the domain wall.
Moreover, when $\Delta_s(x)$ is additionally interpolated such as $\Delta_s(x)=\Delta_{s,0}\tanh{(x/\xi)}$, a new type of domain-wall state emerges connecting $A'$ and $B''$ groundstates [Fig.~\ref{fig.4.GKmodel}(b)].
Without loss of generality, we take the characteristic lengths of the order parameters to be equal.
For comparison, the domain-wall configuration connecting $A'$ and $B'$ is also considered with $\Delta_s = \text{const.}$, which prohibits a zero-energy state because $\Delta_s$ behaves as a Dirac mass term similar to $m_z$ of the RM domain wall. 
The three domain-wall configurations of $A\rightarrow B$, $A'\rightarrow B'$, and $A'\rightarrow B''$ are denoted as Kitaev, constant $s$-wave (CSW), and CI domain walls, respectively.
The energy spectra for these three domain-wall configurations are shown in Fig.~\ref{fig.4.GKmodel}(c); the Kitaev (CSW) domain-wall states are located at zero (non-zero), as expected.
Interestingly, the CI domain wall exhibits the emergence of unexpected Majorana zero-energy domain-wall states.
To compare the domain-wall states with the usual edge states,
we also investigate the  midgap states localized at the left edge for the three cases [Fig.~\ref{fig.4.GKmodel}(c)]; an expected zero-energy Majorana state for the Kitaev domain wall while being split by the Dirac mass for the CSW and CI domain walls.

\begin{figure}[t]
\includegraphics[width=0.9\linewidth]{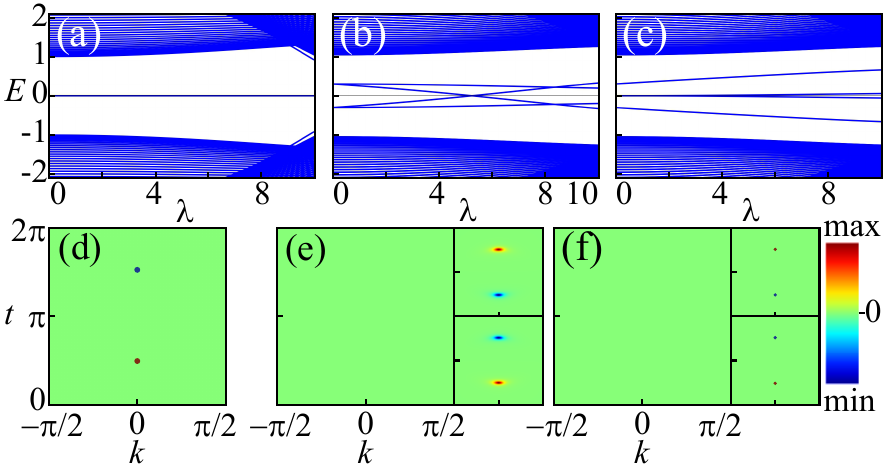}
\caption{\label{fig.5.GKdisorder}
The energy eigenvalues and Berry curvature distributions for the \textbf{(a,d)} Kitaev, \textbf{(b,e)} CSW, and \textbf{(c,f)} CI domain walls under quasi-periodic disorder.
The disorder for electrons at orbital $\gamma$ at site $i$ is $\left(-1\right)^\gamma \lambda \cos\left(2\pi\beta i\right)$ with disorder magnitude $\lambda$ and the inverse golden ratio $\beta$, and vice versa for the holes.
In (a,c), the topological zero energy states are stable for reasonable orders of disorders. In (b,c), the non-zero energy states are localized at the left edge.
For the CSW and CI domain walls, Berry curvature distributions for the subblocks are given on the right.
For \textbf{(e)}, the Berry curvature is blurred, consistent with the non-quantization of the Berry phase. 
The parameters and the domain-wall configurations are the same as in Fig~\ref{fig.4.GKmodel}.
}
\end{figure}

The stability of the Majorana domain-wall states is examined using a similar onsite quasi-periodic disorder used for the GSSH case~\cite{roy2021reentrant,roy2023critical}.
As shown in Fig.~\ref{fig.5.GKdisorder}(a-c), the domain-wall states for the Kitaev and CI cases show robustness against disorder, while that of CSW does not.

Using low-energy effective Dirac-type Hamiltonian of Eq.~(\ref{CIHam}) and the Jackiw-Rebbi method, we analyze the stability of the Majorana zero-energy domain-wall state.
For simplicity, the spatial functions, $\bar{\mu}(x)$ and $\Delta_s(x)$ are denoted as $\bar{\mu}$ and $\Delta_s$ in this paragraph.
Taking the Dirac approximation at $k=0$, the effective Hamiltonian is obtained as $H_\text{eff}^\text{CI}=-\bar{\mu}\tau_z+\Delta_s\tau_y\sigma_y+2\Delta_p k \tau_y\sigma_x$, where $k=-i\partial_x$ in the continuum limit.
The resulting zero-energy domain-wall solutions are $\psi(x)=\mathcal{N}\exp\left[\mp\frac{1}{2\Delta_p}\int^xdx'\sqrt{\Delta_s^2+\bar{\mu}^2}\right]\phi_{1,2}(x)$, with $\phi_1(x)=\left(-\Delta_s\pm\sqrt{\Delta_s^2+\bar{\mu}^2},0,0,\bar{\mu}\right)^T$, and $\phi_2(x)=\left(0,\Delta_s\pm\sqrt{\Delta_s^2+\bar{\mu}^2},\bar{\mu},0\right)^T$ 
for the normalization factor $\mathcal{N}$.
Here, the double signs are in order, with the upper (lower) sign indicating $x>0$ ($x<0$).
The two wavefunctions at zero energy are degenerate and form Majorana pairs in the simplified forms 
$\Psi_{\pm}(x)=\phi_1(x)\pm\frac{\bar{\mu}}{\Delta_s\pm\sqrt{\Delta_s^2+\bar{\mu}^2}}\phi_2(x)$ satisfying the Majorana condition of $\Psi_{\pm}^c = \Psi_{\pm}$.
This form is consistent with the numerical form of the wavefunctions given in Fig.~\ref{fig.4.GKmodel}(b).
Under a SU(4) unitary transformation via unitary operator $U=\exp\left[\frac{i}{2}\left(\frac{\pi}{2}\tau_0+\tan^{-1}{\left(\frac{\Delta_s}{\bar{\mu}}\right)}\tau_x\right)\sigma_y\right]$, the effective Hamiltonian is transformed into that of the two-orbital Kitaev model:
\begin{align*}
{H_\text{eff}^\text{CI}}'
=-\text{sgn}(\bar{\mu})\sqrt{\bar{\mu}^2+\Delta_s^2}\tau_z+2\Delta_pk\tau_y\sigma_z.
\end{align*}
In the viewpoint of low-energy effective theory, the GK model is equivalently transformed into the BDI group and, therefore, can have zero-energy topological modes.
The zero-energy domain-wall solution of ${H_\text{eff}^\text{CI}}'$ highlights the restriction for the emergence of the Majorana domain-wall states; $\bar{\mu}$ and $\Delta_s$ are interpolated such that a so-called topological phase transition point where $\sqrt{\bar{\mu}^2+\Delta_s^2}$ is zero exists, similar to the Kitaev model~\cite{nt3}.

Ascending from the low-energy limit and considering the full Hamiltonian, the Berry phase, and as a result, bulk-boundary correspondence is used to explore the Majorana zero-energy domain-wall states further.
A winding number describes the topological order of a superconducting system~\cite{schnyder2008classification,chiu2016classification}.
Equivalently, the Berry/Zak phase can be applied~\cite{berry1984quantal,Zakphase1989,viyuela2016topological}.
The Berry curvature and phase difference are defined in 2D space-time for the distinct phases in Fig.~\ref{fig.4.GKmodel}(c)~\cite{Supple}.
Figure~\ref{fig.5.GKdisorder}(d) shows the calculated Berry curvature, and the resulting Berry phase difference in Table~S2 is quantized to $\pi$ for the Kitaev domain wall.
In the case of CSW and CI domain walls, the Berry curvatures appear to be zero [Fig.~\ref{fig.5.GKdisorder}(e,f)].
However, $H_\text{CI}(k)$ can be block-diagonalized as
\begin{align*}
\label{BlckHam}
H'_\text{CI}(k)=-\left(2t\cos{k}+\mu\right)\sigma_z+\Delta_s\tau_z\sigma_x+2\Delta_p\sin{k}\tau_z\sigma_y,
\end{align*}
which is grouped into subblocks according to the $\tau_z$ eigenvalues.
As can be seen from the electron-hole picture [Fig.~S3(a)], due to the interacting configuration within the model, there exists a basis-changing unitary operation $(c_A^\dagger,c_B^\dagger,c_A,c_B) \rightarrow (c_A^\dagger,c_B,c_B^\dagger,c_A)$, transforming $H_\text{CI}(k)$ into $H'_\text{CI}(k)$.
The Berry phases of individual subblocks are now examined.
The subblocks for the CI domain wall have a quantized Berry phase difference of $\pi$ in the limit $\frac{t}{\Delta_p}\rightarrow0$.
Regarding the subblocks for the CSW domain wall, the Berry phase difference is not quantized, indicating that the topological zero-energy state is forbidden [Table~S2].
Similar to the GSSH model, Majorana zero-energy domain-wall states can also emerge in seemingly trivial systems via our general bulk-boundary correspondence.


\textit{Conclusion}---So far, the discussion has focused on the generalized SSH and Kitaev models for the AI and CI classes.
Regarding the other three classes (A, AII, C), the methodology and the results thereafter are similar to those of the AI (for A, AII) or CI (for C) classes~\cite{Supple}.
We, therefore, conclude that even systems classified as trivial can possess zero-energy topological domain-wall states.
Physical realization of such systems inevitably results in quantized electric polarization for the GSSH model and Majorana zero modes for the GK models.
Our results are for the simplest model possible, however, there is no loss of generality, and the extensions will be numerous~\cite{note2}.

The realization of our models is expected in various 1D electronic and superconducting systems, such as atomic nanowires~\cite{cheon2015,kim2017}, artificial electronic lattices~\cite{drost2017topological, huda2020tuneable}, graphene nanoribbons~\cite{li2021topological, groning2018engineering}, optical systems~\cite{meier2016, ozawa2019}, proximity-effect-induced superconductors~\cite{fu2008superconducting,nakosai2013two,stanescu2010proximity,guan2016superconducting,chang2016topological}, and hybrid 1D superconductors~\cite{nadj2014observation,
kim2020long,kim2018toward}.
Further expansion could open new platforms in topological science, with potential applications in quantum computation and topological devices, including topological lasers~\cite{stern2013topological, nayak2008, ozawa2019}. 

\begin{acknowledgments}
This work was supported by the National Research Foundation of Korea (NRF) funded by the Ministry of Science and ICT (MSIT), South Korea (Grants No. NRF-2022R1A2C1011646,  NRF-2022M3H3A1085772, No. RS-2023-00252085, and No. RS-2023-00218998).
This work was also supported by the Quantum Simulator Development Project for Materials Innovation through the NRF funded by the MSIT, South Korea (Grant No. NRF-2023M3K5A1094813).
S.H. Han, M. Kang, and S. Cheon also acknowledge support from the POSCO Science Fellowship of the POSCO TJ Park Foundation.
\end{acknowledgments}


\begin{thebibliography}{54}
\bibitem [{par()}]{paraSSH}%
  \BibitemOpen
  \href@noop {} {}\bibinfo {note} {The hopping parameter is $t_0=1.2$ for all
  three cases with dimerization $\Delta$ interpolated from $0.24$ to $-0.24$.
  The number of atoms is $L=2n-1=249$. The characteristic length
  $\xi_1=\xi_2=5$. The onsite energy $m_z$ is fixed at $0.3$ for the RM domain
  wall and interpolated to $-0.3$ for the AI case.}\BibitemShut {Stop}%
\bibitem [{KT()}]{KT}%
  \BibitemOpen
  \href@noop {} {}\bibinfo {note} {For all three cases, the hopping parameter
  and $p$-wave pairings are fixed at $t=1, \Delta_p=10$, respectively, and the
  chemical potential $\mu$ is interpolated from $1$ to $-5$. The characteristic
  length is $\xi=5$, and the number of sites is $L=4n=1600$. The $s$-wave
  pairing term is fixed at $0$, $0.3$ or interpolated from $0.3$ to $-0.3$ for
  the Kitaev, CSW, and CI domain walls, respectively.}\BibitemShut {Stop}%
\bibitem [{not()}]{note2}%
  \BibitemOpen
  \href@noop {} {}\bibinfo {note} {For instance, both spinless and spinful
  models for the A class are suggested in Sec.~S1, and two generalized Kitaev
  models are suggested for the C class in Sec.~S2.}\BibitemShut {Stop}%
\end{thebibliography}

%

\end{document}